\documentclass[final,twoside,11pt,twocolumn,a4]{article}
\usepackage{macro}
\usepackage{pstricks}
\usepackage{graphicx}
\usepackage{amsmath}
\newcommand {\BBB}  {{\cal B}}
      \newcommand {\Om}   {\Omega}
          \newcommand{\figbasicconstruction}{1}
    \newcommand{\fthreepictures}{2}
    \newcommand{\figinvisible}{3}
      \newcommand {\pl}   {\partial}

\begin{document}
\title{\LaTeX Bodies of zero total cross section and bodies invisible in one direction}
\author{\underline{A.\ Aleksenko}$^{1,*}$, A.\ Plakhov$^{1,2}$\\
\textnormal{$^{1}$ Department of Mathematics, Aveiro
University, Aveiro 3810, Portugal\\
$^{2}$ Aberystwyth University, Aberystwyth SY23 3BZ, UK, on leave from Department of Mathematics, University of Aveiro, Aveiro 3810-193, Portugal\\
$^{*}$Email: alena-aleksenko@rambler.ru, axp@aber.ac.uk}}
\date{}
\maketitle

\pagestyle{empty}
\thispagestyle{empty}

\section*{Abstract}
We consider a body in a parallel flow of non-interacting
particles. The interaction of particles with the body is perfectly
elastic.
  We introduce the notions of a body of zero resistance and an invisible
  body   and prove that all such bodies do exist.
\section*{Introduction}

 Suppose that there is
parallel flow of non-interacting particles falling on $\BBB
\subset {\cal R}^3$ which is a bounded connected set with
piecewise smooth boundary. Initially, the velocity of a particle
equals $v_0 \in S^2$; then it makes several reflections from
$\BBB$, and finally moves freely with the velocity $v_\BBB^+(x,
v_0)$, where $x \in \{v_0 \}^\perp$ indicates the initial position
of the particle. One can imagine that the flow is highly rarefied
or consists of light rays.  The force of pressure of the flow on
the body, or {\it resistance of the body in the direction $v_0$,}
is proportional to $R_{v_0}(\BBB) := \int_{\{ v_0 \}^\perp} (v_0 -
v_\BBB^+(x, v_0))\, dx$, where the ratio equals the density of the
flow/medium and $dx$ means the Lebesgue measure in $\{ v_0
\}^\perp$. The {\it problem of minimal resistance} is concerned
with minimizing the resistance in a prescribed class of bodies.
There is a large literature on this problem, starting from the
famous Newton's aerodynamic problem \cite{N}.

We say that the body is {\it invisible in the direction $v_0$} if
the trajectory of each particle outside a prescribed bounded set
coincides with a straight line. We prove that {\it there exist
bodies of zero resistance and bodies invisible in one direction}.

Consider the class of bodies $\BBB$ that are contained in the
cylinder $\Om \times [0,\, h]$ and contain a cross section $\Om
\times \{ c \}$,\, $c \in [0,\, h]$. For the sake of brevity, we
shall call them {\it bodies inscribed in the cylinder}. Multiple
reflections are allowed. If $\Om$ is the unit circle then the
infimum of resistance equals zero, $\inf_\BBB |R_{v_0}(\BBB)| = 0$
(see \cite{P2}). The infimum is not attained, that is, zero
resistance bodies do not exist. This follows from the following
simple proposition.

{\bf Lemma.}\label{predl 1} Let $\Om$ be a convex set with
nonempty interior and let $\BBB$ be a body inscribed in the
cylinder $\Om \times [0,\, h]$. Then $R_{v_0}(\BBB) \ne 0$.

{\bf Proof.}  Using that the particle trajectory does not
intersect the section $\Om \times \{ c \}$ and  $\Om$ is convex,
one concludes that the particle initially moves in the cylinder
above this section, then intersects the lateral surface of the
cylinder and moves freely afterwards. This implies that
$v_\BBB^+(x, v_0) \ne v_0$, hence $R_{v_0}(\BBB) \ne 0$.

\section{Zero resistance bodies and invisible bodies}
{\bf Theorem.}\label{theor 1} There exist (a) a body that has zero
resistance in the direction $v_0$; (b) a body invisible in the
direction $v_0$.

{\bf Proof.} (a) Consider two identical coplanar equilateral
triangles $\rm ABC$ and $\rm A'B'C'$, with $\rm C$ being the
midpoint of the segment $\rm A'B'$, and $\rm C'$, the midpoint of
$\rm AB$. The vertical line $\rm CC'$ is parallel to $v_0$. Let
$\rm A''$ ($\rm B''$) be the point of intersection of segments
$\rm AC$ and $\rm A'C'$ ($\rm BC$ and $\rm B'C'$, respectively);
see Fig.\,{\figbasicconstruction}.
\begin{figure}[h]
\begin{picture}(0,220)
\scalebox{0.9}{ \rput(4.7,4){
\pspolygon[linestyle=dashed,linewidth=0.3pt](-4,-3.4641)(4,-3.4641)(0,3.4641)
\pspolygon[linestyle=dashed,linewidth=0.3pt](-4,3.4641)(4,3.4641)(0,-3.4641)
 \psline[linecolor=black,arrows=->,arrowscale=1.8](-4,4.5)(-4,3.6)
\psline[linecolor=black,arrows=->,arrowscale=1.8](-4,4.5)(-4,3.4641)(1.88,0.069282)
\psline[linecolor=black,arrows=->,arrowscale=1.8](-4,3.4641)(2,0)(2,-4.25)
 \psline[linecolor=black,arrows=->,arrowscale=1.8](-2,4.5)(-2,0.15)
\psline[linecolor=black,arrows=->,arrowscale=1.8](-2,4.5)(-2,0)(4,-3.4641)(4,-4.2)
 \psline[linecolor=black,arrows=->,arrowscale=1.8](-3.5,4.5)(-3.5,2.75)
\psline[linecolor=black,arrows=->,arrowscale=1.8](-3.5,4.5)(-3.5,2.598075)(1.6,-0.34641)
\psline[linecolor=black,arrows=->,arrowscale=1.8](-3.5,2.598075)(2.5,-0.866025)(2.5,-4.25)
 \psline[arrows=<->,arrowscale=1,linewidth=0.6pt](-4,4.3)(-3.5,4.3)
 \rput(-3.73,4.52){\small $dx$}
  \psline[arrows=<->,arrowscale=1,linewidth=0.6pt](2,-3.59)(2.5,-3.59)
 \rput(2.27,-3.77){\small $dx$}
\pspolygon[fillstyle=solid,fillcolor=lightgray](-4,-3.4641)(-4,3.4641)(-2,0)
\pspolygon[fillstyle=solid,fillcolor=lightgray](4,-3.4641)(4,3.4641)(2,0)
\rput(-4.25,-3.4641){A} \rput(-4.3,3.3){A$'$}
  \rput(-3.65,2.43){\small E}
\rput(4.27,-3.3){B} \rput(4.25,3.464){B$'$} \rput(0,3.75){C}
\rput(0,-3.75){C$'$} \rput(-2.4,0){\small A$''$}
\rput(2.4,0){\small B$''$}
   \rput(2.7,-0.8){\small F}
 \psline[arrows=->,arrowscale=1.8](-5,4.5)(-5,3.46)
 \rput(-5.35,4){$v_0$}
} }
\end{picture}
\label{fig basic construction} \caption{The basic construction.}
\end{figure}
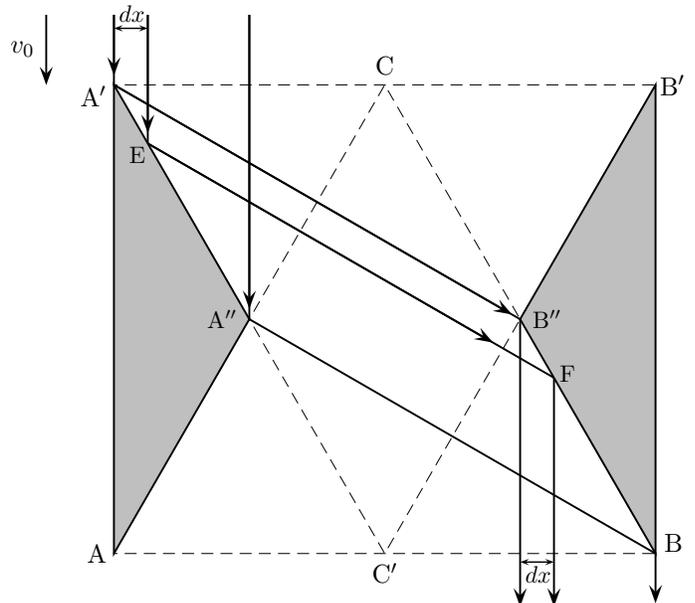
The body $\BBB$ generated by rotation of the triangle $\rm AA'A''$
(or $\rm BB'B''$) around the axis $\rm CC'$ is shown on
Fig.\,{\fthreepictures}a. It has zero resistance in the direction
$v_0$.
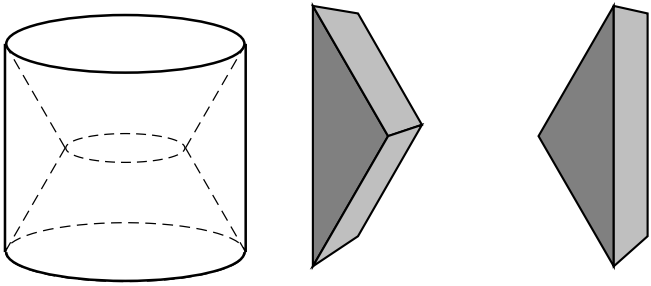
\begin{figure}[h]
\begin{picture}(0,130)
  \scalebox{0.8}{
  \rput(2.1,2.3){
   \psellipse[linewidth=1.2pt](0,-1.73205)(2,0.5)
   \psframe[linecolor=white,fillstyle=solid,fillcolor=white,linewidth=0.6pt](-2,-1.73205)(2,-1.2)
      \psellipse[linestyle=dashed,linewidth=0.6pt](0,-1.73205)(2,0.5)
         \psellipse[linewidth=1.2pt](0,1.73205)(2,0.5)
 \psline[linewidth=1.2pt](-2,-1.73205)(-2,1.73205)
 \psline[linestyle=dashed,linewidth=0.6pt](-2,-1.73205)(-1,0)
 \psline[linestyle=dashed,linewidth=0.6pt](-1,0)(-2,1.73205)
     \psellipse[linestyle=dashed,linewidth=0.6pt](0,0)(1,0.25)
  \psline[linewidth=1.2pt](2,-1.73205)(2,1.73205)
  \psline[linestyle=dashed,linewidth=0.6pt](2,-1.73205)(1,0)
  \psline[linestyle=dashed,linewidth=0.6pt](1,0)(2,1.73205)
}}
 \scalebox{0.5}{ \rput(12.0,4){
\pspolygon[fillstyle=solid,fillcolor=gray,linewidth=1.6pt](-4,-3.4641)(-4,3.4641)(-2,0)
   \pspolygon[fillstyle=solid,fillcolor=lightgray,linewidth=1.6pt](-4,3.4641)(-2,0)(-1.1,0.3)(-2.8,3.2641)
   \pspolygon[fillstyle=solid,fillcolor=lightgray,linewidth=1.6pt](-4,-3.4641)(-2,0)(-1.1,0.3)(-2.8,-2.6641)
\pspolygon[fillstyle=solid,fillcolor=gray,linewidth=1.6pt](4,-3.4641)(4,3.4641)(2,0)
   \pspolygon[fillstyle=solid,fillcolor=lightgray,linewidth=1.6pt](4,-3.4641)(4,3.4641)(4.9,3.2641)(4.9,-2.6641)
} }

\end{picture}
\label{f 3 pictures} \caption{Bodies of zero resistance.}
\end{figure}

 Let the particle first hit the segment ${\rm
A}'{\rm A}''$ at a point E. (If the particle first hits ${\rm
B}'{\rm B}''$, the argument is the same.) After the reflection,
the direction of motion forms the angle $\pi/3$ with the vertical.
Next, the particle hits the segment ${\rm B}''{\rm B}$ at the
point F such that $|{\rm A}'{\rm E}| = |{\rm B}''{\rm F}|$, and
after the second reflection moves vertically downward. That is,
the final velocity equals $v_0$.

(b) A body invisible in the direction $v_0$ can be obtained by
doubling a zero resistance body; see Fig.\,{\figinvisible}.
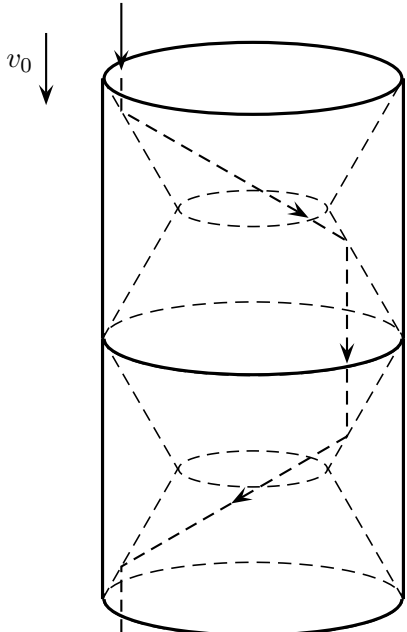
\begin{figure}[h]
\begin{picture}(0,230)
\scalebox{1}{ \rput(5,2.3){
   \psellipse[linewidth=1.2pt](0,-1.73205)(2,0.5)
   \psframe[linecolor=white,fillstyle=solid,fillcolor=white,linewidth=0.6pt](-2,-1.73205)(2,-1.2)
      \psellipse[linestyle=dashed,linewidth=0.6pt](0,-1.73205)(2,0.5)
         \psellipse[linewidth=1.2pt](0,1.73205)(2,0.5)
   \psframe[linecolor=white,fillstyle=solid,fillcolor=white](-2,1.73205)(2,2.3)
      \psellipse[linestyle=dashed,linewidth=0.6pt](0,1.73205)(2,0.5)
 \psline[linewidth=1.2pt](-2,-1.73205)(-2,1.73205)
 \psline[linestyle=dashed,linewidth=0.6pt](-2,-1.73205)(-1,0)(-2,1.73205)
   \psellipse[linewidth=1.2pt](0,5.19615)(2,0.5)
     \psellipse[linestyle=dashed,linewidth=0.6pt](0,3.4641)(1,0.25)
     \psellipse[linestyle=dashed,linewidth=0.6pt](0,0)(1,0.25)
  \psline[linewidth=1.2pt](2,-1.73205)(2,1.73205)
  \psline[linestyle=dashed,linewidth=0.6pt](2,-1.73205)(1,0)(2,1.73205)
     \psline[linewidth=1.2pt](-2,1.73205)(-2,5.19615)
     \psline[linestyle=dashed,linewidth=0.6pt](-2,1.73205)(-1,3.4641)(-2,5.19615)
 \psline[linewidth=1.2pt](2,1.73205)(2,5.19615)
 \pspolygon[linestyle=dashed,linewidth=0.6pt](2,1.73205)(1,3.4641)(2,5.19615)
 \psline[linecolor=black,arrows=->,arrowscale=1.8](-1.75,6.2)(-1.75,5.3)
\psline[linecolor=black,linestyle=dashed,arrows=->,arrowscale=1.8](-1.75,5.2)(-1.75,4.763138)(0.74,3.34)
\psline[linecolor=black,linestyle=dashed,arrows=->,arrowscale=1.8](0.8,3.3)(1.25,3.0310875)(1.25,1.4)
 \psline[linecolor=black,linestyle=dashed,arrows=->,arrowscale=1.8](1.25,1.33)(1.25,0.4330125)(-0.29,-0.46)
 \psline[linecolor=black,linestyle=dashed](-0.29,-0.46)(-1.75,-1.299038)(-1.75,-1.9)
 \psline[linecolor=black,arrows=->,arrowscale=1.8](-1.75,-2)(-1.75,-2.7)
 \psline[arrows=->,arrowscale=1.8](-2.75,5.8)(-2.75,4.8391)
 \rput(-3.1,5.4){$v_0$}
} }
\end{picture}
\label{fig invisible} \caption{A body invisible in the direction
$v_0$. It is obtained by taking 4 truncated cones out of the
cylinder.}
\end{figure}

Denote by $m = m(\BBB, v_0)$ the maximal number of reflections of
an individual particle from the body.

{\bf Theorem}\label{predl 2} (a) If the body $\BBB$ has zero
resistance  in the direction $v_0$ then $m(\BBB, v_0) \ge 2$. (b)
If $\BBB$ is invisible in the direction $v_0$ then $m(\BBB, v_0)
\ge 4$. These inequalities are sharp: there exist zero resistance
bodies  and invisible bodies with exactly 4 reflections.

{\bf Proof.} (a) If $m = 1$ then the final velocity of each
particle does not coincide with the initial one, $v^+_\BBB(x, v_0)
\ne v_0$, therefore $R_{v_0}(\BBB) \ne 0$. That is, a zero
resistance body requires at least two reflections.

(b) Note that a thin parallel beam of particles changes the
orientation under each reflection. To be more precise, let $x(t) =
x + v_0 t$,\, $v(t) = v_0$ be the initial motion of a particle,
and let $x(t) = x^{(i)}(x) + v^{(i)}(x) t$,\, $v(t) = v^{(i)}(x)$
be its motion between the $i$th and $(i+1)$th reflections, $i =
0,\, 1, \ldots, m$. Let the body be invisible in the direction
$v_0$; then one has $v^{(0)} = v^{(m)} = v_0$,\, $x^{(0)} = x$,
and $x^{(m)} - x \perp v_0$. At each reflection and for any fixed
$x$, the orientation of the triple $\big( \frac{\pl x^{(i)}}{\pl
x_1},\,  \frac{\pl x^{(i)}}{\pl x_2},\, v^{(i)} \big)$ changes.
The initial and final orientations, $\big( \frac{\pl x^{(0)}}{\pl
x_1},\,  \frac{\pl x^{(0)}}{\pl x_2},\, v^{(0)} \big)$ and $\big(
\frac{\pl x^{(m)}}{\pl x_1},\,  \frac{\pl x^{(m)}}{\pl x_2},\,
v^{(m)} \big)$, coincide, therefore $m$ is even.

One easily sees that with $m=2$, the initial and final parts of a
trajectory cannot belong to a straight line; this proves that
$m=4$.

\section*{Acknowledgements}
This work was supported by {\it Centre for Research on
Optimization and Control} (CEOC) from the ''{\it Funda\c{c}\~{a}o
para a Ci\^{e}ncia e a Tecnologia}'' (FCT), cofinanced by the
European Community Fund FEDER/POCTI, and by FCT: research project
PTDC/MAT/72840/2006.


\begin{thebibliography}{99}

\bibitem{N}
I. Newton. \textit{Philosophiae Naturalis Principia Mathematica},
1686.

\bibitem{P2}
A.\,Yu. Plakhov. \textit{Newton's problem of the body of minimal
resistance with a bounded number of collisions}. Russ. Math. Surv.
{\bf 58}, 191-192 (2003).

\bibitem{alpl}
A. Aleksenko, A. Plakhov, \textit{Bodies of zero resistance and
bodies invisible in one direction}, 2008,    arXiv:0809.0108v3
[math.DS]
\end{thebibliography}
\end{document}